\documentclass[aps,pra,twocolumn,twoside,superscriptaddress]{revtex4}

\usepackage{bbm}
\usepackage{amsmath}
\usepackage{color}
\usepackage{graphicx,epsfig}

\renewcommand{\d}{{\textrm{d}}}
\newcommand{\tr}{{\textrm{tr}}}

\begin{document}

\title{Thermal state entanglement in harmonic lattices}

\author{Janet Anders\footnote{email: janet@qipc.org}}
\affiliation{	Centre for Quantum Technologies, National University of Singapore, 3 Science Drive 2, Singapore 117543.}
\affiliation{Department of Physics and Astronomy, University College London, London WC1E 6BT, United Kingdom.}

\date{2 June 2008}

\begin{abstract}

We investigate the entanglement properties of thermal states of the harmonic lattice in one, two and three dimensions. We establish the value of the critical temperature for entanglement between neighbouring sites and give physical reasons. Further sites are shown to be entangled only due to boundary effects. Other forms of entanglement are addressed in the second part of the paper by using the energy as witness of entanglement. We close with a comprehensive diagram showing the different phases of entanglement versus complete separability and propose techniques to swap and tune entanglement experimentally. 

\end{abstract}


\maketitle

\section{Introduction}

Entanglement is a purely quantum phenomenon essential for understanding how information is shared in quantum systems and implementing new computing technologies such as measurement-based computation. The existence of entanglement in a composite system implies strong correlations between its parts; so strong that no classical model can explain them. When classical correlations become strong in a condensed matter system, i.e.  when they stretch over large distances, a new `phase' emerges which manifests in qualitatively different behaviour of the material. Phase transitions happen when a parameter, such as the temperature, reaches a critical value. In this paper we will determine the \emph{critical temperature} below which entanglement is present and at which a quantum-correlated phase emerges in the harmonic lattice.

The entanglement properties of the bisected one-dimensional harmonic chain have been analysed in  \cite{AEPW} using extensive mathematical formalism to discuss the ground state (see also \cite{Botero}). The authors also provide a remark on the situation for thermal states, however, the treatment there is rather short. In the present paper we illuminate the thermal case in greater detail by first discussing two-site entanglement and then any entanglement for finite size chains as well as in the thermodynamic limit, for varying external potential and in one, two and three dimensions. Our intention is to gain intuition about the physics behind the quantum-classical transition at $T_{\textrm{crit}}$. We explain how the excitation and mixing of the phononic modes degrades the entanglement unless the lattice is in the intrinsically `quantum phase' below $T_{\textrm{crit}}$ where quantum zero-point fluctuations dominate. We find that entanglement depends on external parameters as well as size and propose that manipulating these will enable techniques of tuning and swapping entanglement in trapped ions. 

The outline of the paper is as follows. In Section \ref{sec:model} the harmonic lattice model is introduced and we summarise technical methods that build the fundament for the results discussed later. Readers familiar with these techniques may skip to Section \ref{sec:analytical} where we derive analytical expressions for the entanglement of the one-dimensional harmonic chain in the thermodynamic limit and at zero temperature, and then approximate the critical temperature for nearest neighbour entanglement. In Section \ref{sec:E4TandN} we present comprehensive numerical results for two-site entanglement in the one-dimensional chain. The value of the respective critical temperatures is explained and finite chain sizes are discussed. In Section \ref{sec:del+d} we investigate the dependence of the nearest neighbour entanglement in the one-dimensional chain when increasing an external trapping potential. Finally, this  behaviour is compared with the behaviour of entanglement in the two- and three-dimensional lattice. 

In Section \ref{sec:E=EW} a different approach to detect entanglement is taken. Using the energy as a witness for any entanglement (not just between two sites) we derive the minimal energy bound for completely separable thermal states. We then establish the critical temperature for witnessed entanglement for the physical sites and find that it is independent of the dimension of the lattice. In Section \ref{sec:phasedia} we show and compare the two temperature thresholds derived in the present paper with the threshold for block-entanglement found in \cite{AEPW} and the threshold for full separability derived in \cite{Winter07} in a comprehensive entanglement phase diagram.  We draw concluding remarks in Section \ref{sec:discussion}, propose possible uses of our results in controlling and storing information in trapped ion systems and discuss the notion of an entangled phase. 

\section{The model} \label{sec:model}

\subsection{Hamiltonian and phononic spectrum}

A harmonic lattice can be realised, for example, as a system of ions that are trapped at $N$ sites by a harmonic trapping potential $\delta$. The ions shall couple to their nearest neighbours with equal strength $\omega$ throughout the lattice.
The Hamiltonian for the $d$-dimensional lattice is  
\begin{equation}\label{eq:ham}
	\hat H = \sum_{\vec R} \left( 
		{{\hat{\vec p}}^{\, 2}_{\vec R} \over 2m} + {m \delta^2 {\hat{\vec u}}^{\, 2}_{\vec R} \over 2} 
					\right)
		+ \sum_{\langle \vec R,  \vec R'\rangle}
		 { m \omega^2 (\hat{\vec u}_{\vec R} - \hat{\vec u}_{\vec R'})^2 \over 2}.
\end{equation}
Here $ \hat{\vec u}_{\vec R}$ is the operator for the deviation from the equilibrium position of each ion. The sites are indexed by $\vec R = \vec n \, a$ with $a$ being the lattice constant and the components of $\vec n$, running over $n_j =  1, 2, ..., \sqrt[d]{N}$. Assuming square geometry of the lattice the length in each spatial direction is $L = \sqrt[d]{N} \, a$ and $m$ is the effective mass of the ions. $ \hat{\vec p}_{\vec R}$ is the momentum operator corresponding to $ \hat{\vec u}_{\vec R}$, so that for all components $i$ and $j$ the commutation relations hold $[\hat u^{(i)}_{\vec R}, \hat p^{(j)}_{\vec R'}] = i \hbar \, \delta_{\vec R, \vec R'} \,  \delta_{i,j}$. The first sum in Eq.(\ref{eq:ham}) is taken over all $N$ sites, and the second sum takes all neighbouring pairs of sites,  $\langle  \vec R, \vec  R'\rangle$, that are harmonically coupled. To simplify the calculation we will adopt periodic boundary conditions and additionally require $\sqrt[d]{N}$ to be an odd number for the sake of definitive calculations. 

The Hamiltonian can be diagonalised by applying first a Fourier transformation followed by a Bogoliubov transformation into a new set of position and momentum operators, see for instance \cite{Brink}. The frequencies of the vibrational modes of the lattice, the phonons, are given by
\begin{equation} \label{eq:omegal}
	\omega_{\vec l} =  2 \omega \sqrt{
	  \sin^2 {\pi l_1 \over \sqrt[d]{N}} 
	+ ...
	+\sin^2 {\pi l_d \over \sqrt[d]{N}} 
	+ \left( {\delta \over 2 \omega} \right)^2},
\end{equation}
with $l_j = - {\sqrt[d]{N} -1 \over 2}, ..., {\sqrt[d]{N} -1 \over 2}$ for a $d$-dimensional square lattice with $N$ sites, each having $2 \,d$ neighbours. It will turn out that the frequency spectrum of the phononic modes is key to the value of the critical temperature for nearest neighbour entanglement and the entanglement properties can be modulated by varying the interaction and the trapping potential. (In the following vector arrows indicating the dimension will be dropped and the dimension of the lattice will be stated explicitly when  necessary.)

\subsection{Thermal states and covariance matrix} 

In this work we are interested in the states of thermal equilibrium, $\rho_{\beta} = { 1\over Z} \,  e^{-\beta \hat H}$, at inverse temperature $\beta = 1/k_B T$. For the harmonic lattice these can be written as a tensor product over all phononic modes, 
\begin{equation}
	\rho_{\beta} = \bigotimes_l \, {e^{-\beta \hbar \omega_{l} \, \hat n_{l}}
					 \over \left(1 - e^{-\beta \hbar \omega_{l}} \right)^{-1}},
\end{equation}
where $\hat n_{l}$ are the number operators for each phononic mode.  
Since the Hamiltonian is quadratic the thermal states are Gaussian states (see for example \cite{contvarstates, Anders-dipl} and references therein) which are specified uniquely by their first and second moments, i.e.  the  expectation values of the canonical operators themselves  and the expectation values of any combination of two canonical operators. 
The first moments can be changed by choice of the coordinate-system, i.e. by applying local, single mode displacements, and their absolute value is of no great physical significance. Only their relative values obey conservation laws. Correlations between modes are determined by higher moments and for Gaussian states the second moments are hence vital for the discussion of entanglement. The second moments of a state $\rho$ with modes 1, 2, .... and canonical operators $ \hat Q \equiv  (\hat u_1, \hat p_1, \hat u_2, \hat p_2, ...)$ are collected in the \emph{covariance matrix}, whose elements are given by 
\begin{equation}\label{eq:CM}
	\Gamma_{jk}(\rho) =   \langle \hat Q_{j} \hat Q_{k} 
	+ \hat Q_{k} \hat Q_{j} \rangle_{\rho}
	- 2 \langle \hat Q_{j} \rangle_{\rho} \langle \hat Q_{k}\rangle_{\rho}.
\end{equation}
$\Gamma$ is a real, symmetric and positive matrix which reveals significant properties of the state such as the occurrence and amount of entanglement in the state. The advantage of formulating the discussion in terms of the covariance matrix is the substantial reduction of parameters, $2\times$ (number of modes), compared to the $2^{(\textrm{number of modes})}$ for the density matrix. 

For the harmonic lattice the covariance matrix for two sites $R$ and $P$ can be found using the transformation from real space into phononic modes in which the Hamiltonian is diagonal. Then expectation values are easily calculated and one obtains
\begin{equation} \label{eq:CMRP}
	\Gamma_{RP} (\rho_{\beta}) = 
	\left[\begin{array}{cccc}
		A &  0 &  E & 0\\
		0 &  B &  0 &  F \\
		E & 0 &  A &  0 \\
		0 & F &  0 &  B\\
	\end{array} \right],
\end{equation}
where the entries are given by 
(for a one-dimensional chain, similar expressions hold for higher dimensions)
\begin{eqnarray} \label{eq:entries1}
	\begin{aligned}
	A = 2 \langle \hat u_{ R} \hat u_{ R} \rangle
		=  {\hbar \over  N m \omega}  \sum_{l}  
			{\coth \xi x_{l}  \over x_{l}},  \\
	B = 2 \langle \hat p_{ R} \hat p_{ R} \rangle 
		=  {m \hbar \omega \over N}  \sum_{l}    
			x_{l} \, \coth \xi x_{l},
	\end{aligned}
\end{eqnarray} 
which are independent of $R$ because of translational symmetry, and 
\begin{eqnarray} \label{eq:entries2}
	\begin{aligned}
	E =   \langle \hat u_{ R} \hat u_{ P} \rangle + \mbox{c.c.}
		=  {\hbar \over  N m \omega}  \sum_{l}  
			\cos \left({2 \pi  l  \over N} r \right)\,  
			{\coth \xi x_{l}  \over x_{l}}, \\
	F = \langle \hat p_{ R} \hat p_{ P} \rangle + \mbox{c.c.}
		=  {m \hbar \omega \over  N}  \sum_{l} 
			\cos \left({2 \pi  l  \over N} r \right)\,  
		 	x_{l} \, \coth \xi x_{l},
	\end{aligned}
\end{eqnarray}
which only depend on the distance between the two sites measured in terms of the lattice constant $a$, $r = {|R - P| \over a} \in \{1, 2, 3, ..., {N-1 \over 2}\}$. Here the inverse temperature is replaced by a unit-free quantity $\xi = {\beta \hbar \omega \over 2} $ and the phononic frequencies are re-scaled by the interaction strength of nearest neighbours $ x_l = {\omega_l \over \omega}$.

\subsection{Two-site entanglement condition}

For Gaussian continuous variable states of two modes there exists a necessary and sufficient criterion  that decides whether the modes are entangled. (The criterion is necessary and sufficient for Gaussian states of $1 \times N$ modes, \cite{Werner:Wolf}.) It is a variation of the positive partial transpose criterion (PPT-criterion) for discrete systems \cite{PPTcrit,HHH96}. For separability of two sites $R$ and $P$ in the lattice the criterion requires the positivity of a matrix inequality,   
\begin{equation} \label{eq:PPT}
	\rho_{RP} \mbox{ separable} 
	\Leftrightarrow \Gamma_{RP}^{T_P} (\rho) + i \hbar  \bigoplus_{R,P} \sigma \ge 0,
\end{equation}
where $\sigma = \left[\begin{array}{cc} 0 & 1 \\ -1 & 0 \end{array} \right]$ is the symplectic form. $\Gamma_{RP}^{T_P}$ is the covariance matrix with time-reversed mode $P$,
\begin{equation}
	\Gamma_{RP}^{T_P} = \left( \mathbbm{1}  \oplus \left[\begin{array}{cc} 1 & 0 \\0 & -1 \end{array} \right] \right)\, 
 			\Gamma_{RP} \,
			\left(\mathbbm{1}  \oplus  \left[\begin{array}{cc} 1 & 0 \\ 0 & -1 \end{array} \right] \right),
\end{equation}
i.e. for $\Gamma_{RP}$ in Eq.~\eqref{eq:CMRP} we simply replace $F$ by $-F$ to obtain $\Gamma_{RP}^{T_P}$. Using the symmetry of $\Gamma_{RP}^{T_P}$ criterion \eqref{eq:PPT} reduces to requiring the positivity of two temperature dependent functions, $S_{1,2} \, (T, r)$, 
\begin{eqnarray} \label{eq:separability}
	 0 \le    S_{1,2} \, (T, r) &=& {1 \over \hbar^2} \, 
	     			\langle \left( \hat u_R \pm \hat u_P\right)^2\rangle
				\langle \left( \hat p_R \mp \hat p_P\right)^2\rangle -1, \nonumber \\
	    			&=& {1 \over \hbar^2} \, (A \pm E(r))(B \mp F(r)) -1.
\end{eqnarray}
Here $r = {|R-P| / a}$  is the separation of the two sites $R$ and $P$ in the one-dimensional chain. If positivity fails  the two sites are entangled and the magnitude of their entanglement can be given by the logarithmic negativity $E_{\mathcal{N}}$, defined in \cite{Vidal02, JensPhD}.  $E_{\mathcal{N}}$ can be calculated directly from the functions $S_{1,2}$ as 
\begin{equation}\label{eq:logneg}
	E_{\mathcal{N}}   = \sum_{j=1}^2 \, \max \{0, - \ln \sqrt{S_j + 1}\}.
\end{equation}

Analytical and numerical results based on criterion \eqref{eq:PPT} for two sites are discussed in Section \ref{sec:analytical} and Section \ref{sec:E4TandN}, respectively. The criterion is also sufficient to detect entanglement between any two \emph{blocks} of sites as shown in \cite{AEPW}. Another approach to reveal other kinds of entanglement, i.e. exclude full separability, is the method of entanglement witnesses, introduced in \cite{HHH96, Terhal02}. Entanglement witnesses are observables that can take a particular range of expectation values only for entangled state and therefore `witness' entanglement. They have been used to investigate spins systems \cite{Brukner, Toth, Dowling} and also \cite{Wu}, as well as Bosonic gases \cite{Anders06}. In Section \ref{sec:E=EW} we use the same strategy for the harmonic lattice to deduce a temperature below which entanglement must be present. A third approach to investigate entanglement in a harmonic lattice uses the full separability condition for the covariance matrix, $\Gamma \ge \bigoplus_j \Gamma_j$, for all sites $j$ \cite{Werner:Wolf}. The mathematical derivation following this direction is given in \cite{Winter07}. Fig.~\ref{fig:exact-phase} in the present paper unites the results on nearest neighbour entanglement and witnessed entanglement with these previous works and shows how the different strategies complete each other.
 
\section{Two-site entanglement (1D)} \label{sec:analytical}

\subsection{Entanglement in the ground state} \label{subsec:T01D}

Any pair of nearest neighbours in the harmonic chain is coupled and we expect it to be entangled in the ground state.  Before going in the discussion of the numerical plots in Section \ref{sec:E4TandN}, let us here derive the negativity Eq. (\ref{eq:logneg}) for zero temperature and in the thermodynamical limit. 

Let $T \to 0$, i.e. $\beta \to \infty$, then 
$S_{1,2} (r) = {1\over N^2} \, \sum_{l, k} 
			{x_k \over x_l}  
			\left(1 \pm  \cos {2 \pi l \over N} \cdot r \right)
			\left(1 \mp \cos {2 \pi k \over N} \cdot r \right) - 1.$ 
Further substituting ${\pi k \over N} = y$ and ${\pi l \over N} = z$ and taking the continuum limit $N \to \infty$ to replace the sums by integrals one obtains
\begin{eqnarray}
S_{1,2} (r) &=& {1 \over \pi^2}  
 		\left( \int_{-{\pi \over 2}}^{{\pi \over 2}} \d z \,
		{1 \pm \cos 2 z r  \over  \sqrt{\sin^2 z + \left( {\delta \over 2 \omega} \right)^2}}
		\right) \\  \times && \nonumber
		\left( \int_{-{\pi \over 2}}^{{\pi \over 2}} \d y \, 
		{\sqrt{\sin^2 y + \left( {\delta \over 2 \omega} \right)^2}}
		\left(1 \mp \cos 2 y r  \right) \right) -1.
\end{eqnarray}
Simplifying this expression further we let $\delta \to 0$ and by using the symmetry of the integrands one obtains 
$S_{1,2}(r)= {4 \over \pi^2}  
 		\left( \int_{0}^{{\pi \over 2}} \d z \, {1 \pm  \cos 2 z r  \over |\sin z| } \right) 
		\left( \int_{0}^{{\pi \over 2}} \d y \, {|\sin y|} \left(1 \mp \cos 2 y r  \right) \right) -1,$ 
which can be easily evaluated for arbitrarily distant neighbours. For nearest neighbours we find
\begin{eqnarray}
S_{1} (r=1) 	&\to& + \infty, \\
S_2	(r=1)		&=& {16 \over 3 \pi^2} -1 \approx - 0.46.
\end{eqnarray}
Finding $S_2 (r=1)$ negative implies that nearest neighbours are entangled and this can be quantified in terms of the negativity, Eq.~\eqref{eq:logneg}, 
\begin{equation} \label{eq:T0negativity}
		E_{\mathcal{N}} (r =1)  	\approx 0.31.
\end{equation}

The behaviour of entanglement between any other pair of two sites in the chain is not obvious. Even though non-nearest neighbours do not interact directly, common nearest neighbours could mediate entanglement. We find that this is \emph{not} the case when the chain size grows arbitrarily, as both $S_1$ and $S_2$ remain positive for $r=2,3, ...$. However, for finite chain sizes, the above calculation is only an approximation and the numerical results show small entanglement between next nearest neighbours for very small chains at low temperatures, see Section \ref{sec:E4TandN} for discussion. 
To conclude, at zero temperature and  in the thermodynamic limit of large $N$ two-site entanglement only exists between nearest neighbours and not between any further separated sites. This result agrees with \cite{Cramer05}, where the authors showed that in gapped harmonic lattice systems the correlation functions decay exponentially in the ground state and thermal states.

\subsection{Critical temperature} 

When the temperature rises, thermal mixing becomes stronger and correlations between the nearest neighbours decrease. We expect intuitively that any entanglement should vanish at large enough temperatures. The transition into the classical regime is marked by a critical temperature which we here want to approximate. The exact temperature for various parameter ranges can be found numerically and is discussed in Sections \ref{sec:E4TandN} and \ref{sec:del+d}.

Again we take the continuum limit $N \to \infty$ and replace the sums by integrals in Eq.(\ref{eq:separability}). Assuming that the trapping potential vanishes, $\delta \to 0$, we obtain 
$S_{1,2} (r) = {4 \over \pi^2} 	\left( \int_{0}^{{\pi \over 2}} \d z \, {\coth \left(2 \xi | \sin z| \right) \over |\sin z| } \, (1 \pm \cos 2 z r) \right) \, \times \left( \int_{0}^{{\pi \over 2}} \d y \,  
		\coth \left(2 \xi |\sin y| \right) { | \sin y| }
		\left(1 \mp \cos 2 y r  \right) \right) -1.$
The integrals have no closed expression and we need to further approximate the $\coth$-function to evaluate $S_{1,2}$. The coupling parameter $\xi = {\beta \hbar \omega \over 2}$, where $\omega$ is the finite interaction strength between neighbours, is temperature dependent. Assuming high temperatures, $T \to \infty \Rightarrow \beta \to 0$, one can expand to first order in $\xi$: $\coth \left(2 \xi | \sin z| \right) \approx {1 \over 2 \xi | \sin z|} + {2 \xi | \sin z| \over 3}$. Thus for nearest neighbours one obtains  
\begin{eqnarray}
	S_{1} (r=1) &\to& + \infty, \\
	S_{2} (r=1) &=& {1 \over 2\xi^2} - {1 \over 2}  < 0 \quad \mbox{when} \quad  \xi > 1.
\end{eqnarray}
Here we have dropped quadratic and higher powers of $\xi$ for consistency. Indeed $S_2$ can become negative and the point where $S_{2}(r=1)$ crosses through zero gives the critical temperature for nearest neighbour entanglement, from $\xi_{\textrm{n.n.}}= 1$ follows
\begin{equation} \label{eq:Tnn}
 	T_{\textrm{n.n.}} = {\hbar \omega \over 2 k_B}. 
\end{equation}
We remark that this temperature is not as large as the initial approximation assumed and one may wonder about its validity. Higher expansions could be considered to evaluate the critical value more accurately. We will not follow this direction here, instead we will present numerical data \emph{confirming} the derived critical value in the next section.

For the next-nearest neighbours  the factors change slightly and the expressions become $S_{1}(r=2) \to \infty$ and $S_{2}(r=2) = {1 \over \xi^2}$. Both expressions are never negative irrespective of the temperature and no  entanglement is expected between next-nearest neighbours in the continuum limit. The numerical data in following section independently confirm this result.

\section{Entanglement for finite temperature and size (1D)} \label{sec:E4TandN}
 
\begin{figure}[tb]	
	\begin{flushleft}
	\quad$E_{\mathcal{N}}$\\[1ex]
	{\includegraphics[width=0.43\textwidth]{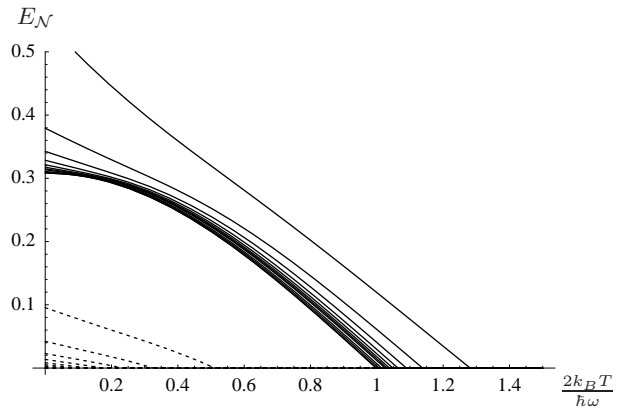}$\!\!\! {2 k_B T \over \hbar \omega}$}
	\caption{\label{fig:nnE+nnnE}  
The figure shows the negativity $E_{\mathcal{N}}$ of a one-dimensional chain in thermal equilibrium over the temperature $T$ (expressed in multiples of the coupling $\omega$).
The trapping potential is small compared to the coupling, $\delta = 10^{-4} \omega$.
The solid curves show the nearest neighbour entanglement for an increasing number of sites: $N = 3$ (uppermost), $..., 51$ which converge to an asymptotic curve for $N \to \infty$ (lowest of the solid curves).
The dashed curves show the next-nearest neighbour entanglement for an increasing number of sites, $N = 3$ (uppermost), $..., 15$, which vanishes for further increasing $N$. }
	\end{flushleft}
\end{figure}

\subsection{Data} 

Numerical data of nearest neighbour (solid) and next-nearest neighbour (dashed) entanglement in terms of the logarithmic negativity $E_{\mathcal{N}}$, Eq.(\ref{eq:logneg}) are displayed in Fig.~\ref{fig:nnE+nnnE} over the temperature and for varying size of a one-dimensional chain. We observe that in general the entanglement is highest for low temperatures and decreases sharply with increasing $T$ until it vanishes abruptly at some critical temperature. For nearest neighbours the critical temperature decreases for increasing $N$ and converges to $T_{\textrm{n.n.}} \approx {\hbar \omega \over 2 k_B}$ in the thermodynamical limit, which agrees with the analytical result in Eq.(\ref{eq:Tnn}). Hence, the stronger the interaction $\omega$ between nearest neighbours, the more robust their entanglement with respect to thermal noise. The maximal amount of entanglement at $T=0$ decreases for growing $N$ and converges to the threshold value of  $E_{\mathcal{N}} \approx  0.3$ independent of $\omega$, in good agreement with the analytical result Eq.~\eqref{eq:T0negativity}. 

Next-nearest neighbour entanglement  (dashed curves) is in general much smaller than the nearest neighbour entanglement. Nonetheless, the numerical data show that a small amount of entanglement exists between next-nearest neighbours when the chain is very small. However, the entanglement as well as the critical temperature are considerably lower than for nearest neighbours and decrease rapidly to zero for increasing $N$.

\subsection{Interpretation} 

We can understand the behaviour of entanglement with changing temperature in the following way. Without quantum zero-point fluctuations, the ions in the chain would be completely fixed and entanglement would stretch over all sites at zero temperature. However, due to the fluctuation of the phononic modes of the chain (\# of phononic modes is $N$ with $N \to \infty$) correlations are mixed with anti-correlations and the entanglement in the chain reduces considerably. For nearest neighbours the mixing of vibrations is biased. This is a result of the direct coupling between them leading to an average deviation of the ions from equilibrium position of less than the harmonic oscillator length of the interaction. Indeed for small temperatures entanglement exists between nearest neighbours, see the solid curves in Fig.~\ref{fig:nnE+nnnE}. 

Next-nearest neighbours and more distant sites do not have a bias of the phononic modes in the thermodynamic limit and entanglement between them is averaged out even at $T=0$. However, when the chain size is small, finite size effects occur. So few phononic modes are available that the different swinging motions of the ions are not completely balanced and some correlations survive. This is why we find next-nearest neighbour entanglement for very small lattices, see the dashed curves in Fig.~\ref{fig:nnE+nnnE}. It is also the reason why the entanglement between nearest neighbours is in general slightly higher for smaller lattices, as can be seen for the solid set of curves where the uppermost curve displays the behaviour of a tiny chain with $N=3$, and the lower curves apply for increasing $N$. 

When the temperature is raised, more phonons are excited and the mixing of different vibrations is enhanced. This leads to a rapid decay of the entanglement and when the temperature reaches the magnitude of the phononic modes that drive nearest neighbours against each other the entanglement vanishes completely. This happens when the two degrees of freedom, the position and the momentum of each of the $N$ ions, with classical thermal energy ${k_B T \over 2}$ per degree of freedom exceeds the energy of the quantum fluctuations of each of the $N$ springs that connect nearest neighbours, ${\hbar \omega \over 2}$ (for $\delta \approx 0$). 
We thus expect nearest neighbours to behave classically when $2 \cdot {k_B T \over 2} \ge {\hbar \omega \over 2}$ is fulfilled. This heuristic argument gives the critical temperature for entanglement, $T_{\textrm{n.n.}} = {\hbar \omega \over 2 k_B}$, in agreement with the numerical cutting point in Fig.~\ref{fig:nnE+nnnE}. It also indicates that the energy is a good witness of entanglement, which we will elaborate in Section \ref{sec:E=EW}.

\section{Dependency on frequency spectrum and dimension of space} \label{sec:del+d}

\begin{figure}[tb]
	\begin{flushleft}
	\quad$E_{\mathcal{N}}$\\[1ex]
	{\includegraphics[width=0.43\textwidth]{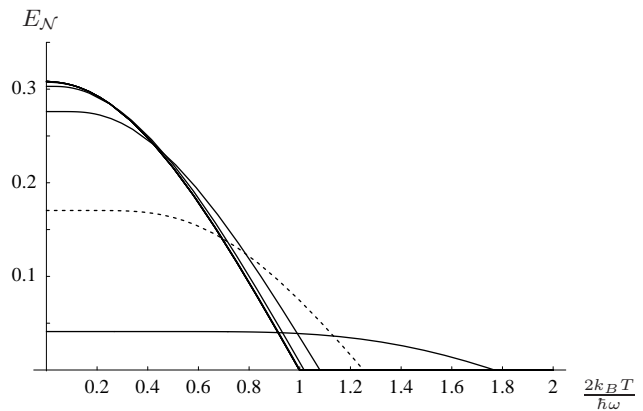}$\, {2 k_B T \over \hbar \omega}$}
    	\caption{\label{fig:N=49} Entanglement, $E_{\mathcal{N}}$, of nearest neighbours in a one-dimensional chain with $N = 49$ sites, depicted over temperature, $T$. The different curves represent increasing ratios of the trapping potential versus the coupling strength, with the uppermost curve reaching $E_{\mathcal{N}} \approx 0.3$ at $T=0$ for $\delta / \omega = 10^{-4}$, and lower curves for the parameters $\delta / \omega =10^{-7/2}, ..., 10^{-1/2}, 1$ (dashed), $\sqrt{10}$ . }
	\end{flushleft}
\end{figure}

\subsection{Data} 

The impact of the phononic frequency spectrum on the entanglement properties are studied quantitatively by varying the ratio between trapping potential and nearest neighbour coupling. The numerical results are displayed in Fig.~\ref{fig:N=49}. As before, 
for small $\delta/\omega$ the entanglement starts at $E_{\mathcal{N}} \approx 0.3$ for $T=0$ and decreases for increasing temperature until it vanishes at $T_{\textrm{n.n.}} (\delta/\omega << 1) \approx {\hbar \omega \over 2 k_B}$. However, when the trapping potential is raised and reaches the magnitude of the interaction the initial entanglement starts off much lower, $E_{\mathcal{N}} \approx 0.17$ for $\delta = \omega$. Notable is that this entanglement is more robust and survives increasingly high temperature mixing. For example, when $\delta = \omega$, the critical temperature has increased to $T_{\textrm{n.n.}} (\delta =\omega) \approx 1.25 {\hbar \omega \over 2 k_B}$.

\subsection{Interpretation} 

When the trapping potential $\delta$ is raised close to and above $\omega$ two effects emerge. Firstly, the on-site potential fixes the ions tightly to their sites and only a  small fraction of the available energy can be put into the interaction.  Even though they are coupled, the ions cannot follow the motion of their neighbours and entanglement between nearest neighbours starts off at substantially lower values for $T=0$ compared to the case of small $\delta$, see Fig.~\ref{fig:N=49}. The second effect is that higher phononic modes `cost'  a higher amount of energy. Thus for growing $\delta$ the bias of phononic modes cannot be compensated even at higher temperatures and nearest neighbours stay entangled up to increasing critical temperatures, as shown in Fig.~\ref{fig:N=49}.

\begin{figure}[tb]
	\begin{flushleft}
	\quad$E_{\mathcal{N}}$\\[1ex]
	{\includegraphics[width=0.43\textwidth]{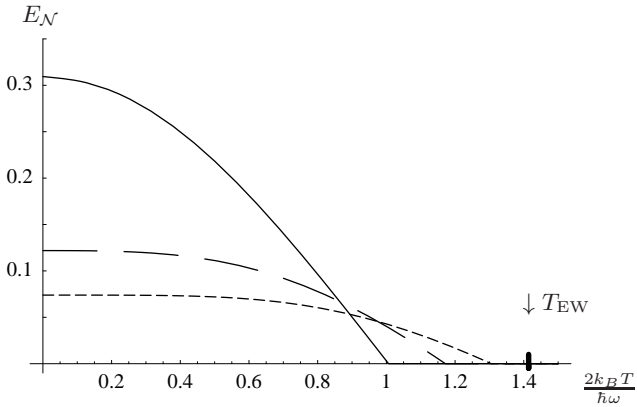}$\, {2 k_B T \over \hbar \omega}$\\[-1.5cm]
	\hspace{0.39\textwidth}$\downarrow T_{\textrm{EW}}$\\[1.2cm]}
    	\caption{\label{fig:alld} Entanglement of nearest neighbours in 1D (solid line, $N=31$), 2D (dashed, $N =31^2$ ) and 3D (points, $N=31^3$) is shown over the temperature. The trapping potential is $\delta / \omega = 10^{-4}$ in all cases. For a discussion please refer to the main text. The dimension-independent temperature for any witnessed entanglement, $T_{\textrm{EW}}$, derived in Section \ref{sec:E=EW} is also indicated and discussed in Section \ref{sec:phasedia}. }
	\end{flushleft}
\end{figure}

For higher dimensions, that is, when the chain becomes a two- or three-dimensional square lattice, the behaviour of two-site entanglement is qualitatively similar to the one-dimensional case, as shown in Fig.~\ref{fig:alld}. Comparing this figure with the graphs for varying $\delta$ in Fig.~\ref{fig:N=49} one can see that raising the dimension leads to similar effects as raising the trapping potential. The higher the dimension the lower is the ground state entanglement, but the higher rises the critical temperature. 

The reason behind this behaviour is the growing fixation of the ions to their sites for increasing dimension. Namely, in the thermodynamic limit (and for $\delta \to 0$) the average displacement of the ions from their equilibrium position,  $\langle {\hat{\vec u}_{\vec R}}^2 \rangle$, diverges linearly in 1D and logarithmically in 2D, while converging in 3D. No long-range order can be established in 1D and 2D as the crystal simply `melts' (for a discussion see for example \cite{Brink} and references therein). Only in three dimensions do the fluctuations of the ionic positions become smaller than the lattice spacing, long-range order can occur and a solid with a three-dimensional crystalline structure can form. To obtain the same effect one could impose a strong trapping potential to hold the ions tightly in place and thereby force crystallisation in low dimensions. Thus, increasing the trapping potential is effectively raising the dimension of the lattice structure and vice versa.

\section{Energy as an entanglement witness in $d$ dimensions} \label{sec:E=EW}

Two-site entanglement is a very restricted kind of entanglement. Other kinds of entanglement exist, for example genuine three-partite entanglement. As indicated by the previous numerical results the balance between the available thermal energy compared to the phononic spectrum determines the existence of entanglement implying that the energy may be a useful criterion as an entanglement witness. This idea has been used in a number of works before, for example in \cite{Brukner,Dowling,Toth,Anders06}. Adopting this strategy for the harmonic lattice in $d$ dimensions we identify the temperature range where some kind of entanglement must necessarily be present.  

\subsection{Derivation of energy bound for separated sites.} 

The most general, fully separable (into $N$ sites) state is of the form 
\begin{equation} \label{eq:total-separable}
	\rho_{\textrm{sep}} = \sum_{j} ~ p_j ~
		\rho_j^1 \otimes \rho_j^2  \otimes  ... \otimes \rho_j^N,
\end{equation}  
where $p_j \ge 0$ are probabilities with $\sum_j p_j = 1$. 
The energy $\langle \hat H \rangle$ (c.f. Eq.~\eqref{eq:ham}) can be evaluated for all separable states as
\begin{eqnarray}
	\langle \hat H \rangle_{\textrm{sep}} 
	=	\sum_{j, R} p_j ~ \tr\left[\rho_j^R \left(
		{\hat p_R^2 \over 2m} + {m \delta^2 \hat  u_R^2\over 2} 
		+ {2 m \omega^2 \hat u_R^2\over 2}
		\right)\right] \nonumber \\
	- 	 m \omega^2  \sum_{j, \langle R,R' \rangle} p_j ~
		 \tr\left[\rho_j^R \left({\hat u_R}	\right)\right] 
		\tr\left[\rho_j^{R'} \left({\hat u_{R'}}	\right)\right]. \quad
\end{eqnarray}
The second term can w.l.o.g. be set zero since all first order expectation values of any operator can be shifted locally to zero without affecting the non-local entanglement properties. Thus, the energy of the whole lattice in any separable configuration, $\rho_{\textrm{sep}}$, becomes the sum of the individual energies for all sites,
\begin{equation} \label{eq:total=sumofparts}
	\langle \hat H \rangle_{\textrm{sep}}  = \sum_{R} \tr\left[\rho^R ~ \hat H_R \right],
\end{equation}
where $\rho^R = \sum_j ~ p_j ~ \rho_j^R$ is an arbitrary state for site $R$ and $\hat H_R :=  {\hat p_R^2 \over 2m} + {m \delta^2 \hat  u_R^2\over 2} + {2 m \omega^2 \hat u_R^2\over 2}$ the effective Hamilton operator for one site $R$ alone, where the $\hat  u_R$ and $\hat  p_R$ are in general $d$-dimensional operators. Thus the description of the lattice has become a mean-field model where each site interacts with an effective, identical background produced by the interaction with all other sites. Reversely, a description using only effective one-site Hamiltonians leads to a classical model where entanglement has been averaged out and states are always separable. 

$\hat H_R$ can be rewritten as a harmonic oscillator, $\hat H_R = \hbar \Omega \left(\hat n_R + {d \over 2}\right)$, where the frequency $\Omega = \sqrt{2 \omega^2 +\delta^2}$ is identical for all sites $R$. By the presence of quantum zero-point fluctuations we always have $\langle \hat H_R \rangle \ge   {d \over 2}~ \hbar \Omega$ for any state $\rho^R$. The possible energy for a separable configuration of the whole lattice is thus bounded from below by
\begin{equation} \label{eq:bound-sites}
	\langle \hat H \rangle_{\textrm{sep-sites}} 
		\ge  \sum_{R} {d \over 2}~ \hbar \Omega
		= {d \over 2}~N \hbar \Omega.
\end{equation}
Any state having an energy below this bound is necessarily entangled with respect to the sites. 

\subsection{Entanglement for different modes.}

It may seem puzzling that one could apply the same strategy for other set of modes, for instance, for the phononic modes themselves and find a non-zero energy bound. Indeed, for the normal modes one would find
\begin{equation} 
	\langle \hat H \rangle_{\textrm{sep-phonons}} 
		\ge  {d \over 2}~\sum_{\vec l} \, \hbar \omega_{\vec l},
\end{equation}
with $\omega_{\vec l}$ as given in Eq.~\eqref{eq:omegal} for $d$ dimensions. This inequality is \emph{always fulfilled} as the right hand side gives just the ground state energy. The criterion is thus useless here as it can never be violated and is hence unable to detect any entanglement. 

By symmetry we expect that the sites in space are `most' entangled, i.e. give the highest energy bound, while the phononic modes remain separable in any thermal state. The situation for `intermediate modes' is between these extremes. To show this we use again the energy witness argument (assuming one dimension for simplicity). For all orthogonal transformations $\mathcal{O}$ that transform the set of phononic momentum operators, $\hat{\mathcal{P}_l}$, into some new momentum operators $\hat P_k$, such that $\hat P_k= \sum_l \, \mathcal{O}_{kl} \, \hat{\mathcal{P}_l}$ and similarly for the position operators, $\hat U_k = \sum_l \, \mathcal{O}_{kl} \, \hat{\mathcal{U}_l}$, so that the commutation relations remain $[\hat U_k, \hat P_m] = i \hbar  \, \delta_{km}$, the Hamiltonian becomes
\begin{eqnarray} 
	\hat H &=& \sum_{l} \left( 
		{\hat{\mathcal{P}}^2_{l} \over 2m} 
		+ {m \omega_l^2 \, \hat{\mathcal{U}}^2_{l} \over 2} 		\right)
		= \sum_{k} \left( 
		{\hat P^2_{k} \over 2m} + {m \mathcal{W}^2_{kk} \, \hat U^2_{k} \over 2} 		\right) \nonumber\\
		&&+ \sum_{k \not = m} \, {m \mathcal{W}^2_{km} \, \hat U_{k} \, \hat U_{m}  \over 2} 
\end{eqnarray}
where the interactions are given by the elements of the matrix $\mathcal{W}^2_{km} = \sum_l \, \mathcal{O}_{kl} \, \omega_l^2 \, \mathcal{O}^T_{lm}$. With the same argument as before we obtain a minimal energy threshold for all separable states with respect to the new set of modes,
\begin{equation} 
	\langle \hat H \rangle_{\textrm{sep-$k$-modes}} 
		\ge  {1 \over 2}~ \sum_{k} \, \hbar  \sqrt{\mathcal{W}^2_{kk}}.
\end{equation}
In particular, the diagonal frequencies are $\mathcal{W}^2_{kk} = \sum_l \, \mathcal{O}^2_{kl} \, \omega_l^2$ with the doubly-stochastic matrix $\mathcal{O}^2$, $\sum_l \, \mathcal{O}^2_{kl} = 1 = \sum_k \, \mathcal{O}^2_{kl}$. With the Birkhoff-von Neumann theorem $\mathcal{O}^2$ can be written as a convex combination of permutation matrices of the same order and because the square-root is concave we only need to consider the two extremal cases which bound $ \sum_{k} \,  \sqrt{\mathcal{W}^2_{kk}}$ from below and above. The first is a single permutation $\Pi$ with $\mathcal{O}^2_{kl} = \delta_{l,\Pi(k)}$ resulting in a reordering $\mathcal{W}^2_{kk} =  \omega_{\Pi(k)}^2$, and the second is the complete mixture of all possible permutations such that $\mathcal{O}^2_{kl} = {1 \over N}$ and $\mathcal{W}^2_{kk} = {1 \over N} \, \sum_l \, \omega_l^2 = \Omega^2$. These extremes are indeed identical to the phononic modes and the sites, respectively. The energy bound for the general set of modes is thus bounded from below and above,
\begin{equation}
 	 {1 \over 2} \sum_l \, \hbar \omega_l 
		\le  {1 \over 2} \sum_k \, \hbar \sqrt{ \mathcal{W}^2_{kk}} 
			\le {1 \over 2} \, N \, \hbar \Omega,
\end{equation}
being bigger than the zero-point energy, while remaining smaller than the energy bound for individual sites. Site entanglement thus possesses the highest energy bound.

\subsection{Critical temperature for separated sites.} 

The energy of the $d$-dimensional harmonic lattice in a thermal state can be related to the temperature. The internal energy, $U$, can be explicitly calculated and for high temperatures  one finds the Boltzmann's equi-partition law, where each degree of freedom of the system obtains a thermal energy contribution of ${k_B T \over 2}$,
\begin{equation} \label{eq:UTlarge}
	U (T) \to d ~ N {k_B T} \quad \mbox{for}\quad T \to \infty.
\end{equation}
(In this context the relation is identical to the \emph{Dulong-Petit} law of constant heat capacity at high temperatures.) Setting the internal energy equal to the energy bound for the sites, Eq.~\eqref{eq:bound-sites}, one finds that any fully separable configuration has a temperature above a certain critical value $T_{\textrm{EW}}$ 
\begin{equation}  \label{eq:TEW2}
	T_{\textrm{sep}} \ge {\hbar \Omega \over 2 k_B} 
		=  {\hbar \omega \over 2 k_B}  \sqrt{2 + {\delta^2 \over \omega^2}} 
		\equiv T_{\textrm{EW}}.
 \end{equation}
Conversely, any configuration with a lower temperature must necessarily contain entanglement between the sites of the lattice. 
Remarkably, the critical value is independent of the \emph{dimension} of the system. Furthermore, the witness temperature for $\delta = 0$, $T_{\textrm{EW}} = {1 \over \sqrt{2}}\, {\hbar \omega \over k_B} \approx 1.4142 \, {\hbar \omega \over 2 k_B}$ is close to the nearest neighbour critical temperature $T_{\textrm{n.n.}} \approx {\hbar \omega \over 2 k_B}$ in 1D. In higher dimensions the nearest neighbour critical temperature increases and approaches the witness-temperature, as shown in Fig.~\ref{fig:alld}. 

\medskip

The energy relation  Eq.~\eqref{eq:UTlarge} and hence the witness temperature Eq.~\eqref{eq:TEW2} is initially only valid in the asymptotic limit of high temperatures. For small temperatures the behaviour of the internal energy $U(T)$ and its derivative, the heat capacity $C_V(T)$, has been subject of intense discussion at the beginning of the 20th century (which is covered in many textbooks, see e.g. \cite{heatcapacitydebate}). Introducing a heuristic \emph{cut-off frequency} $\omega_{D}$, Debye found that in three dimensions the phononic oscillations lead to a heat capacity proportional to $T^3$, implying $U_{D} \propto T^4$ for temperatures below $T_D={\hbar \omega_D \over k_B}$. With the same method the internal energy scales with $U_{D} \propto T^{d+1}$ in lower dimensions. Discussing the Debye temperature in detail is beyond the scope of this paper, however, direct conclusions can be drawn just from the properties of the energy function. Because of the monotonicity of the internal energy and its convexity at small temperatures, $U_{D} (T)$  is below the value of the high-temperature limit, $U(T)$ in Eq.~\eqref{eq:UTlarge}, in the regime $T < T_D$. The witness temperature derived with the high temperature approximation, $T_{EW}$, is thus a lower bound to the critical temperature that one would obtain when using the low temperature formula. The entangled region detected with the high temperature approximation is thus a subset of the region which one could detect when using the small temperature approximation. This implies that $T_{EW}$ is a valid bound for witnessed entanglement also at small temperatures, being most tight in one dimension and lesser so in higher dimension due to the stronger bend of the $T^{d+1}$-law.

\section{Entanglement phase diagram} \label{sec:phasedia}

\begin{figure}[tb]  
	\centering
		 \includegraphics[width=0.48\textwidth]{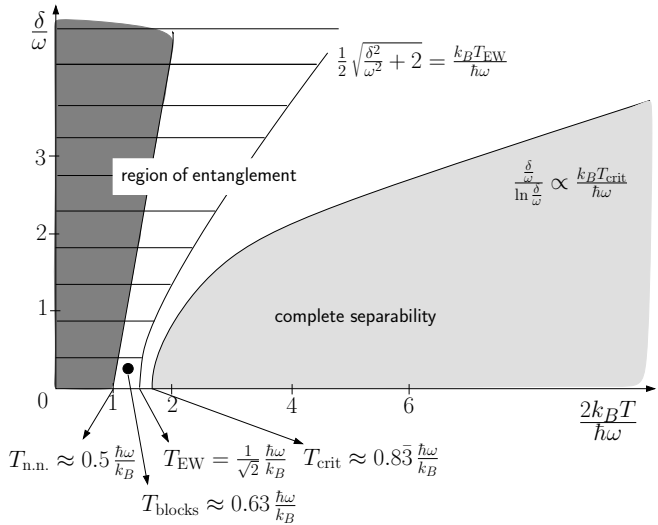} 
		 \caption{\label{fig:exact-phase} 
		 The graphic keeps the exact scale relations of the different phases of entanglement and separability in the harmonic chain ($N=49$), for varying trapping potential $\delta$ and over the temperature $T$. 
		 Nearest neighbour entanglement exists in the dark shaded region up to the temperature threshold $T_{\textrm{n.n.}}$.
		 Entanglement is witnessed by the energy in the hatched area, with threshold temperature $T_{\textrm{EW}}$.
		 Additionally the temperature threshold, $T_{\textrm{blocks}}$ (black point), for the existence of entanglement between two neighbouring blocks of size $N/2$  from \cite{AEPW} (for a single value of $\delta$) is displayed. 
		The chain passes into the light shaded region of complete separability at the temperature $T_{\textrm{crit}}$ from \cite{Winter07}. 
		For $\delta= 0$ the threshold values are displayed below the figure, for their discussion please refer to the text. }
\end{figure}

Fig.~\ref{fig:exact-phase} is a comprehensive and exact diagram comparing the two derived temperature thresholds for nearest neighbour entanglement and witnessed entanglement with previous results from \cite{AEPW} and \cite{Winter07}. As found in Section \ref{sec:E4TandN}, two-site entanglement resides between nearest neighbours (dark shaded) for temperatures below  $T_{\textrm{n.n.}} \approx 0.5 \, {\hbar \omega / k_B}$ (for $N \to \infty$, $\delta =0$) while non-nearest neighbour entanglement vanishes for large $N$. The hatched set are states whose non-separability is detected by the energy witness. It fully includes all nearest neighbour entanglement and extends further up to the temperature $T_{\textrm{EW}} \approx {1 \over \sqrt{2}} \, \hbar \omega /k_B$, cf. Eq.~\eqref{eq:TEW2}, where no two-site entanglement exists. A different phase of how entanglement is distributed is thus reached. 

For example, entanglement can exist between two neighbouring \emph{blocks} of $N/2$ sites. Results in \cite{AEPW} for $\delta = \omega / \sqrt{20}$ (this appears to be the critical value according to Figure 12 in~\cite{AEPW} after retaining all units) confirm this and their critical temperature for block-entanglement, $T_{\textrm{blocks}} \approx 0.63 \, \hbar \omega /k_B$, is indicated in Fig.~\ref{fig:exact-phase} by the dark point. When the temperature is raised further the chain passes over into complete separability (light shaded), i.e. no entanglement exists for any possible split. The temperature at which this transition happens is given as $T_{\textrm{crit}} = \hbar \omega /1.2 \, k_B \approx 0.8\bar 3 \, \hbar \omega /k_B$ (for $\delta = 0$ ) in \cite{Winter07}. For sufficiently small $\delta$, the witness bound $T_{EW}$ is in good agreement with the exact threshold for full separability, $T_{\textrm{crit}}$ \cite{Winter07}. This indicates that the temperature range where the entanglement survives thermal mixing is identical to the regime where the quantum zero-point effects are important. The use of the energy as an entanglement witness is thus justified and in general an alternative, easier way to evaluate regions where entanglement is present, c.f. \cite{Brukner, Toth, Dowling, Anders06}.
For higher dimension we find that the critical temperature for nearest neighbour entanglement increases towards the dimension independent witness temperature (see Fig.~\ref{fig:alld}) thus leaving less temperature range where non-two-site entanglement is witnessed.

When $\delta$ is large compared to $\omega$, the frequency spectrum shrinks to essentially one dominating frequency, $\omega_l \approx \delta$. The mixing of frequencies occurring at increasing temperatures becomes  irrelevant and entanglement persist up to higher temperatures, the higher $\delta$ becomes. The asymptotic behaviour of the witness-temperature becomes proportional to $\delta$ as $\frac{k_B T_{\textrm{EW}}}{\hbar \omega} \propto \frac{\delta}{2 \omega}$. Yet the full-separability temperature scales with $\delta \over \ln \delta$ as $\frac{k_B T_{0}}{\hbar \omega} \sim \frac{\delta/\omega}{2\ln(\delta/\omega)}$ which implies a growing gap between the witnessed and the true entangled region for growing $\delta$. 

Interestingly in this model any entanglement must be non-PPT \cite{Cavalcanti07}, in other words, there always exists a \emph{bi-partite split} of the chain into two parties such, that partial transposition on one of the parties results in a non-positive density matrix. Any entanglement can thus potentially be identified by this property and the value of the negative eigenvalue can be used to measure the amount of entanglement.

\section{Conclusion and Outlook}  \label{sec:discussion}

In condensed matter physics the two-point correlation function \cite{correlation-function} between two sites is an important quantity used to determine the behaviour of the material. If correlations stretch over a wide range of sites, i.e. the correlation function becomes a constant or decreases polynomially for increasing distances, then one speaks of the establishment of an ordered phase.
Quantum physics allows for a stronger kind of correlation, that is, entanglement. The entanglement criterion Eq.~\eqref{eq:separability} is reminiscent of the classical order criterion for the two-point correlation function. Instead of requiring that the fluctuations of the ionic positions, $\langle \hat u^2_R\rangle$, should be smaller than the lattice spacing, we here have a condition on the product of the relative positions and the relative momentum fluctuations. In other words, instead of requiring order in \emph{space} the entanglement criterion  \eqref{eq:separability} requires order between two sites in \emph{phase space}.

We have found that entanglement exists between nearest neighbours in all dimensions, for all lattice sizes, and for all values of $\delta$, when the system is cooled below some critical temperature, which depends on the former parameters. By changing the temperature the entanglement can be modulated and even cut out. The critical temperature at which the entanglement vanishes is directly proportional to the coupling constant; for the harmonic chain this is the coupling $\omega$. This relation is in agreement with previous results in spin chains \cite{Brukner} and Bose gases \cite{Anders06,Heaney07} where also the interaction strength determines the critical temperature. 

Entanglement between next-nearest neighbours exists in very small lattices, but is in general much weaker in magnitude and also only observable at much lower temperatures than for the nearest neighbour case. A possible application for trapped ions in lattices with harmonic interaction, would be to tune the value of $\delta$ and $T$ such, that only nearest neighbour ions remain entangled. By slowly inserting an additional ion in the lattice one can destroy entanglement to the former nearest neighbours and the new nearest neighbours become entangled.  Varying the trapping potential $\delta$ or the interaction strength $\omega$, one can further change the amount and range of entanglement in the lattice. Such techniques and the possibility to apply additional external potentials that modify the interaction and hence affect the structure of the entanglement may allow to store and manipulate information on the ions.  For instance, applying an extra harmonic potential to all ions in a small periodic chain results in a higher frequency of the lowest phononic mode, the centre of mass mode, that could influence the chain to `swap' entanglement from nearest neighbours to opposite ions in the ring by simply raising the temperature. 

Even though in the thermodynamic limit only neighbouring pairs are entangled and more distant pairs are not, we can speak of an `entangled phase'. This notion is justified because entanglement is much stronger than classical correlations and nearest neighbour entanglement is already sufficient to transport information across the whole lattice. Such correlations enable universal quantum computation in the measurement-based model, as initiated in \cite{Raussendorf01} for spin systems with nearest neighbour interaction and its analogue for the continuous variable case proposed in \cite{Menicucci06}. 
The entangled phase survives, approximately, as long as the system is in the quantum regime where the zero-point fluctuations dominate the physics of the lattice. Going beyond that range implies the validity of a mean-field theory where each site can be described as an independent oscillator whose effective frequency  includes the contributions from the rest of the system.

\acknowledgments

J.A. is supported by the Gottlieb Daimler und Karl Benz-Stiftung. This work was supported in part by the Singapore A*STAR Temasek Grant No. 012-104-0040 and this research is part of QIP IRC www.qipirc.org (GR/S82176/01). I wish to thank A. Ekert, B.-G. Englert, L. C. Kwek, Ch. Miniatura, J. Suzuki, V. Vedral, A. Winter for many insightful discussions on the subject.

\vfill




\begin{thebibliography}{99}
\bibitem{correlation-function}
The two-point correlation function $G(\vec R)$ measures the correlation between the order-parameter at two positions separated by $\vec R$.  A solid is characterised by long-range order of the mass density $\rho(x)$. For the harmonic lattice the two-point correlation function can be expressed as $G(\vec R) = \sum_{\vec k} e^{i \vec k \cdot \vec R} \sum_{\vec P} e^{i \vec k \cdot \vec P} \langle e^{i \vec k \cdot (\hat u_{\vec P} - \hat u_{\vec 0})} \rangle$, which can be calculated explicitly, see for example \cite{Brink, Chaiken}.
\end{thebibliography}
\end{document}